\documentclass[aps,pra,showpacs,floatfix,reprint]{revtex4-1}
\usepackage{bm}
\usepackage{graphicx} 
\usepackage{amsmath}
\usepackage{amsfonts}
\begin{document}
\title{Optimized Multichannel Quantum Defect Theory for cold molecular collisions}

\author{James F. E. Croft}
\author{Jeremy M. Hutson}
\affiliation{Joint Quantum Centre (JQC) Durham/Newcastle, Department of
Chemistry, Durham University, South Road, Durham, DH1~3LE, United Kingdom}

\author{Paul S. Julienne}
\affiliation{Joint Quantum Institute, NIST and the University of Maryland,
Gaithersburg, Maryland 20899-8423, USA}

\date{\today}

\begin{abstract}
Multichannel quantum defect theory (MQDT) can provide an efficient alternative
to full coupled-channel calculations for low-energy molecular collisions.
However, the efficiency relies on interpolation of the $\bm{Y}$ matrix that
encapsulates the short-range dynamics, and there are poles in $\bm{Y}$ that may
prevent interpolation over the range of energies of interest for cold molecular
collisions. We show how the phases of the MQDT reference functions may be
chosen so as to remove such poles from the vicinity of a reference energy and
dramatically increase the range of interpolation. For the test case of Mg+NH,
the resulting optimized $\bm{Y}$ matrix may be interpolated smoothly over an
energy range of several Kelvin and a magnetic field range of over 1000~G.
Calculations at additional energies and fields can then be performed at a
computational cost that is proportional to the number of channels $N$ and not
to $N^3$.
\end{abstract}

\maketitle

\section{Introduction}
\label{intro}

Samples of cold and ultracold molecules have unique properties that are likely
to have applications in many diverse areas. These include high-precision
measurement \cite{Hudson:2002, Bethlem:2009}, quantum information processing
\cite{DeMille:2002} and quantum simulation \cite{Carr:NJPintro:2009}. There is
also great interest in the development of controlled ultracold chemistry
\cite{Krems:PCCP:2008}.

Atomic and molecular interactions and collisions are crucial to the production
and properties of cold and ultracold molecules. However, quantum-mechanical
molecular collision calculations can be computationally extremely expensive.
Such calculations are usually carried out using the coupled-channel method, in
which the wavefunction is expanded
\begin{equation}
\Psi(r,\tau) = r^{-1}\sum_{i=1}^N \varphi_i(\tau)\psi_{i}(r).
\end{equation}
Here the $N$ functions $\varphi_i(\tau)$ form a basis set for the motion in all
coordinates, $\tau$, except the intermolecular distance, $r$, and $\psi_{i}(r)$
is the radial wavefunction in channel $i$. Substituting this expansion into the
time-independent Schr\"odinger equation and projecting onto the basis function
$\varphi_j(\tau)$ yields a set of $N$ coupled differential equations. The
properties of completed collisions are describe by the scattering matrix $\bm
S$, which is obtained by matching the functions $\psi_i(r)$ to free-particle
wavefunctions (Ricatti-Bessel functions) at long range \cite{Johnson:1973}. In
the full coupled-channel method, explicit solution of the coupled equations
takes a time proportional to $N^3$.

The problems encountered in cold molecular collisions often require very large
number of channels. Atom-molecule and molecule-molecule interaction potentials
can be strongly anisotropic, requiring large basis sets of rotational functions
for convergence. In addition, calculations are often required in an applied
field, where the total angular momentum $J$ is no longer a good quantum number.
Because of this, the large sets of coupled equations cannot be factorized into
smaller blocks for each $J$ as is possible in field-free scattering
\cite{Arthurs:1960}. Furthermore, at the very low collision energies of
interest, small splittings between molecular energy levels have important
consequences. Effects such as tunneling \cite{Zuchowski:2009} and nuclear
hyperfine splitting \cite{Lara:PRA:2007, Gonzalez-Martinez:hyperfine:2011} each
multiply the number of channels.

In cold collision studies, the scattering $\bm S$ matrix is often a fast
function of collision energy $E$ and magnetic field $B$, with extensive
structure due to scattering resonances and discontinuous behavior at threshold.
Calculations are often required over a fine grid of energies and/or applied
electric and magnetic fields, and this further multiplies the computational
expense.

We have recently shown \cite{Croft:MQDT:2011} that Multichannel Quantum Defect
Theory (MQDT) \cite{Seaton:QDT:1966, Seaton:1983, Greene:1982, Mies:1984,
Mies:MQDT:2000, Raoult:2004} provides an attractive alternative to full
coupled-channel calculations in these circumstances. MQDT attempts to represent
the scattering properties in terms of a matrix $\bm Y(E,B)$ \cite{Greene:1982,
Mies:1984, Mies:MQDT:2000, Raoult:2004} that is a smooth function of $E$ and
$B$. If this can be achieved, the matrix can be obtained once and then used for
calculations over a wide range of energies and fields, or obtained by
interpolation from a few points. Once the matrix $\bm{Y}(E,B)$ has been
obtained, the time required for calculations at additional energies and fields
is only proportional to $N$, not $N^3$.

One problem with MQDT is that the $\bm Y$ matrix may have poles as a function
of $E$ and $B$, and these limit the range over which it can be interpolated. In
cold molecular collision studies, calculations are typically needed over an
energy range of order 1~K above threshold, and for magnetic fields up to a few
thousand gauss \footnote{Units of gauss rather than tesla, the accepted SI unit
of magnetic field, are used in this paper to conform to the conventional usage
of this field.}. This contrasts with the situation for collisions of ultracold
atoms, where the energy range of interest is commonly a few $\mu$K and the
fields are typically a few hundred gauss.

In the present paper, we show how MQDT $\bm Y$ matrices can be defined to allow
smooth interpolation over substantial ranges of collision energy and applied
field. This will allow the use of MQDT to provide substantial savings in
computer time.

\section{Theory}

A full description of MQDT has been given previously \cite{Seaton:QDT:1966,
Seaton:1983, Greene:1982, Mies:1984, Mies:MQDT:2000, Raoult:2004}. We give here
only a brief description, following ref.\ \cite{Croft:MQDT:2011}, which is
sufficient to describe the notation we use.

MQDT defines the matrix $\bm{Y}(E,B)$ at a matching distance $r_{\rm match}$ at
relatively short range. The $N$-channel scattering problem at energy $E$ is
partitioned into $N_{\rm o}$ open channels (with $E_i^\infty \leq E$, where
$E_i^\infty$ is the threshold of channel $i$), $N_{\rm c}$ weakly closed
channels, and $N_{\rm s}$ strongly closed channels. Strongly closed channels
are those that make no significant contribution to the scattering dynamics at
$r > r_{\rm match}$.

The scattering dynamics beyond $r_{\rm match}$ is accounted for using
single-channel (uncoupled) calculations in a basis set that diagonalizes the
Hamiltonian at $r=\infty$. The solution of the multichannel Schr\"odinger
equation at $r>r_{\rm match}$ is written in the matrix form
\begin{equation}\label{eqn:match_mqdt}
 \bm{\Psi} = \bm{f}(r) + \bm{g}(r)\bm{Y},
\end{equation}
where $\bm{f}$ and $\bm{g}$ are diagonal matrices containing the functions
$f_i$ and $g_i$, which are linearly independent solutions of a reference
Schr\"odinger equation in each asymptotic channel $i$,
\begin{equation}
\left[-\frac{\hbar^2}{2\mu}\frac{d^2}{dr^2} + U_i^{\rm ref}(r) - E\right] f_i(r) =0,
\label{eq:SEref}
\end{equation}
and similarly for $g_i(r)$. The reference potentials $U_i^{\rm ref}(r)$
approach the true potential at long range, and $\mu$ is the reduced mass. They
include the centrifugal terms $\hbar^2 L_i(L_i+1)/2\mu r^2$, where $L_i$ is the
partial-wave quantum number for channel $i$. $\bm Y$ is an $N_{\rm{ref}}\times
N_{\rm{ref}}$ matrix, where $N_{\rm{ref}}=N_{\rm o}+N_{\rm c}$.

In our approach \cite{Croft:MQDT:2011}, $\bm Y$ is obtained numerically by
matching the solutions of the coupled-channel equations to $f_i(r)$ and
$g_i(r)$ at $r_{\rm match}$. The $\bm{S}$ matrix is then obtained from $\bm{Y}$
using Eqs.\ (21) to (23) of ref.\ \cite{Croft:MQDT:2011}, which require 3 QDT
parameters $C_i$, $\tan\lambda_i$ and $\xi_i$ in each open channel and a single
QDT parameter $\tan\nu_i$ in each weakly closed channel. In the open channels
the reference functions are asymptotically related to Ricatti-Bessel functions
$J_{L_i}(r)$ and $N_{L_i}(r)$ \cite{Johnson:1973},
\begin{equation}
\label{eqn:openQDT}
\begin{pmatrix} f_i \\ g_i \end{pmatrix} =
\left( \begin{array}{cc} C_i & 0 \\ -C_i\tan\lambda_i & C_i^{-1} \end{array} \right)
\left( \begin{array}{rr} \cos\xi_i & \sin\xi_i \\ -\sin\xi_i & \cos\xi_i \end{array} \right)
\begin{pmatrix} J_{L_i} \\ N_{L_i} \end{pmatrix}.
\end{equation}
Here $\xi_i$ is the asymptotic phase shift of the function $f_i$ with respect
to the Ricatti-Bessel function $J_{L_i}$. The QDT parameter $C_i$ relates the
amplitudes of the energy-normalized functions at long range to functions with
Wentzel-Kramers-Brillouin (WKB) normalization at short range, while
$\tan\lambda_i$ describes the modification of the WKB phase due to threshold
effects. Far from threshold, $C_i \approx 1$ and $\tan \lambda_i \approx 0$. In
the weakly closed channels the reference functions are asymptotically
\begin{equation}
\label{eqn:closedQDT}
\begin{pmatrix} f_i \\ g_i \end{pmatrix} =
\left( \begin{array}{rr} \cos\nu_i & \sin\nu_i \\ -\sin\nu_i & \cos\nu_i \end{array} \right)
\begin{pmatrix} \phi_i \\ \gamma_i \end{pmatrix},
\end{equation}
where $\phi_i$ is the solution of (\ref{eq:SEref}) that decays exponentially at
large $r$ and $\gamma_i$ is its linearly independent partner, which is
exponentially growing.

The absolute phases chosen for the reference functions $f_i$ and $g_i$ are
arbitrary, and different choices produce different $\bm{Y}$ matrices and
different MQDT parameters. In particular, Eq.\ (\ref{eqn:match_mqdt}) shows
that a pole in $\bm Y$ occurs whenever the propagated multichannel wavefunction
in any channel $i$ has no contribution from the reference function $f_i$.
However, all phase choices produce the same physical $\bm S$ matrix. We are
therefore free to choose the phase in order to produce a $\bm Y$ matrix with
advantageous characteristics. Here we show how the phase may be chosen to
produce a $\bm Y$ matrix that is pole-free over a wide range of energy or
magnetic field and can be interpolated smoothly.

Rotating the reference functions $f_i$ and $g_i$ by an angle $\theta_i$
gives a new set of linearly independent reference functions $\bar f_i$
and $\bar g_i$,
\begin{equation}
\label{eqn:rotation}
\begin{pmatrix} \bar f_i \\ \bar g_i \end{pmatrix}
= \left( \begin{array}{rr}  \cos\theta_i & -\sin\theta_i \\
\sin\theta_i & \cos\theta_i \end{array} \right)
\begin{pmatrix} f_i \\ g_i \end{pmatrix}.
\end{equation}
These rotated reference functions define a new $\bm Y$ matrix and a new set of
QDT parameters ($\bar C$, $\tan\bar\lambda$, $\bar \xi$ and $\tan\bar\nu$).
Combining equations (\ref{eqn:openQDT}), (\ref{eqn:closedQDT}) and
(\ref{eqn:rotation}) gives
\begin{widetext}
\begin{equation}
\label{eqn:xibar}
\bar{\xi}_i = \arctan\left[\frac{C_i^2 \sin\xi_i (\cos\theta_i+\tan\lambda_i \sin\theta_i)-\cos\xi_i \sin\theta_i}{C_i^2 \cos\xi_i (\cos\theta_i+\tan\lambda_i \sin\theta_i)+\sin\xi_i\sin\theta_i}\right],
\end{equation}
\begin{equation}
\label{eqn:tlbar}
\tan\bar{\lambda}_i=
-\frac{2 C^4_i \tan\lambda_i \cos2\theta_i+
\left[1+C^4_i \left(\tan^2\lambda_i-1\right)\right] \sin2\theta_i}
{2\left(C^4_i \cos^2\theta_i+\sin\theta_i
\left[\sin\theta_i+C^4_i \tan\lambda_i (2 \cos\theta_i+\tan\lambda_i \sin\theta_i)\right]\right)},
\end{equation}
\begin{equation}
\label{eqn:cbar}
\bar{C}_i =
\left(\frac{\sin\xi_i \sin\theta_i}{C_i}+C_i \cos\xi_i (\cos\theta_i+\tan\lambda_i \sin\theta_i)\right)\sqrt{1+\frac{\left(\cos\xi_i \sin\theta_i-C_i^2 \sin\xi_i (\cos\theta_i+\tan\lambda_i \sin\theta_i)\right)^2}{\left(\sin\xi_i\sin\theta_i+C_i^2 \cos\xi_i(\cos\theta_i+\tan\lambda_i \sin\theta_i)\right)^2}},
\end{equation}
\begin{equation}
\label{eqn:nubar}
\bar{\nu}_i = \nu_i-\theta_i.
\end{equation}
\end{widetext}
Far from threshold ($E \gg 1$~K), Eqs.\ (\ref{eqn:xibar}) to (\ref{eqn:nubar})
simplify to $\bar \xi_i = \xi_i - \theta_i$, $\tan \lambda_i \approx 0$,  $C_i
\approx 1$ and $\bar\nu_i = \nu_i - \theta_i$. However, in the threshold region
that is of interest in cold molecule studies, Eqs.\ (\ref{eqn:xibar}) to
(\ref{eqn:nubar}) must be evaluated explicitly.

\subsection{Basis sets and quantum numbers}

As a test case, we consider cold collisions between NH ($^3\Sigma^-$) and Mg
atoms \cite{Wallis:MgNH:2009}. This is the same system as considered in ref.\
\cite{Croft:MQDT:2011}, but the present work uses a larger basis set which
introduces more scattering resonances and denser poles in the $\bm Y$ matrix.

The energy levels of NH in a magnetic field are most conveniently described
using Hund's case (b), in which the molecular rotation $n$ couples to the spin
$s$ to produce a total monomer angular momentum $j$. In zero field, each
rotational level $n$ is split into sublevels labeled by $j$. In a magnetic
field, each sublevel splits further into $2j + 1$ levels labeled by $m_j$, the
projection of $j$ onto the axis defined by the field. For the $n = 0$ levels
that are of most interest for cold molecule studies, there is only a single
zero-field level with $j = 1$ that splits into three components with $m_j =
+1$, $0$ and $-1$.

The coupled equations are constructed in a partly coupled basis set $|nsjm_j
\rangle |LM_L \rangle$, where $L$ is the end-over-end rotational angular
momentum of the Mg atom and the NH molecule about one another and $M_L$ is its
projection on the axis defined by the magnetic field. Hyperfine structure is
neglected. The matrix elements of the total Hamiltonian in this basis set are
given in ref.\ \cite{Gonzalez-Martinez:2007}. The only good quantum numbers
during the collision are the parity $p = (-1)^{n+L+1}$ and the total projection
quantum number $M = m_j + M_L$. The calculations in the present work are
performed for $p=-1$ and $M = 1$. This choice includes s-wave scattering of NH
molecules in initial state $m_j = +1$, which is magnetically trappable, to $m_j
= 0$ and $-1$, which are not. The present work uses a converged basis set
including all functions up to $n_{\text{max}} = 6$ and $L_{\text{max}} = 8$, as
in ref.\ \cite{Wallis:MgNH:2009}.

We label elements of $\bm Y$ and $\bm S$ by subscripts $\alpha,L,M_L
\rightarrow \alpha',L',M_L'$, where $\alpha$ represents an eigenstate of free
NH that may be approximately labeled by $(n,s,j,m_j)$. However, the collisions
considered in the present paper are all among the $n=0, j=1$ levels and so
$\alpha$ is simply abbreviated to $m_j$. For diagonal elements we suppress the
second set of labels.

\subsection{Numerical methods}

The coupled-channel calculations required for both MQDT and the full
coupled-channel approach were carried out using the MOLSCAT package
\cite{molscat:v14}, as modified to handle collisions in magnetic fields
\cite{Gonzalez-Martinez:2007}. The coupled equations were solved numerically
using the hybrid log-derivative propagator of Alexander and Manolopoulos
\cite{Alexander:1987}, which uses a fixed-step-size log-derivative propagator
in the short-range region ($r_\text{min} \le r < r_\text{mid}$) and a
variable-step-size Airy propagator in the long-range region ($r_\text{mid} \le
r \le r_\text{max}$). The full coupled-channel calculations used
$r_\text{min}=2.5$~\AA, $r_\text{mid}=50$~\AA\ and $r_\text{max}=250$~\AA\
(where 1~\AA\ = $10^{-10}$~m). MQDT requires coupled-channel calculations only
from $r_\text{min}$ to $r_{\rm match}$ (which is less than $r_\text{mid}$), so
only the fixed-step-size propagator was used in this case.

The MQDT reference functions and quantum defect parameters were obtained as
described in ref.\ \cite{Croft:MQDT:2011}, using the renormalized Numerov
method \cite{Johnson:1977} to solve the 1-dimension Schr\"odinger equations for
the reference potentials. The MQDT $\bm{Y}$ matrix was then obtained by
matching to the log-derivative matrix extracted from the coupled-channel
propagation at a distance $r_\text{match}$. In this paper all MQDT calculations
use the reference potential
\begin{equation}\label{eqn:refunc3}
U^{\text{ref}}_i(r) =
V_0(r) + \frac{\hbar^2 L_i(L_i+1)}{2\mu r^2} + E_i^\infty,
\end{equation}
where $V_0(r)$ is the isotropic part of the interaction potential. This
reference potential has been shown to produce quantitatively accurate results
when $\bm{Y}$ is reevaluated at each collision energy and magnetic field
\cite{Croft:MQDT:2011}. However, such reevaluation relinquishes most of the
computational savings that MQDT is intended to achieve.

\begin{figure}[tb]
\centering
\includegraphics[width=1\columnwidth]{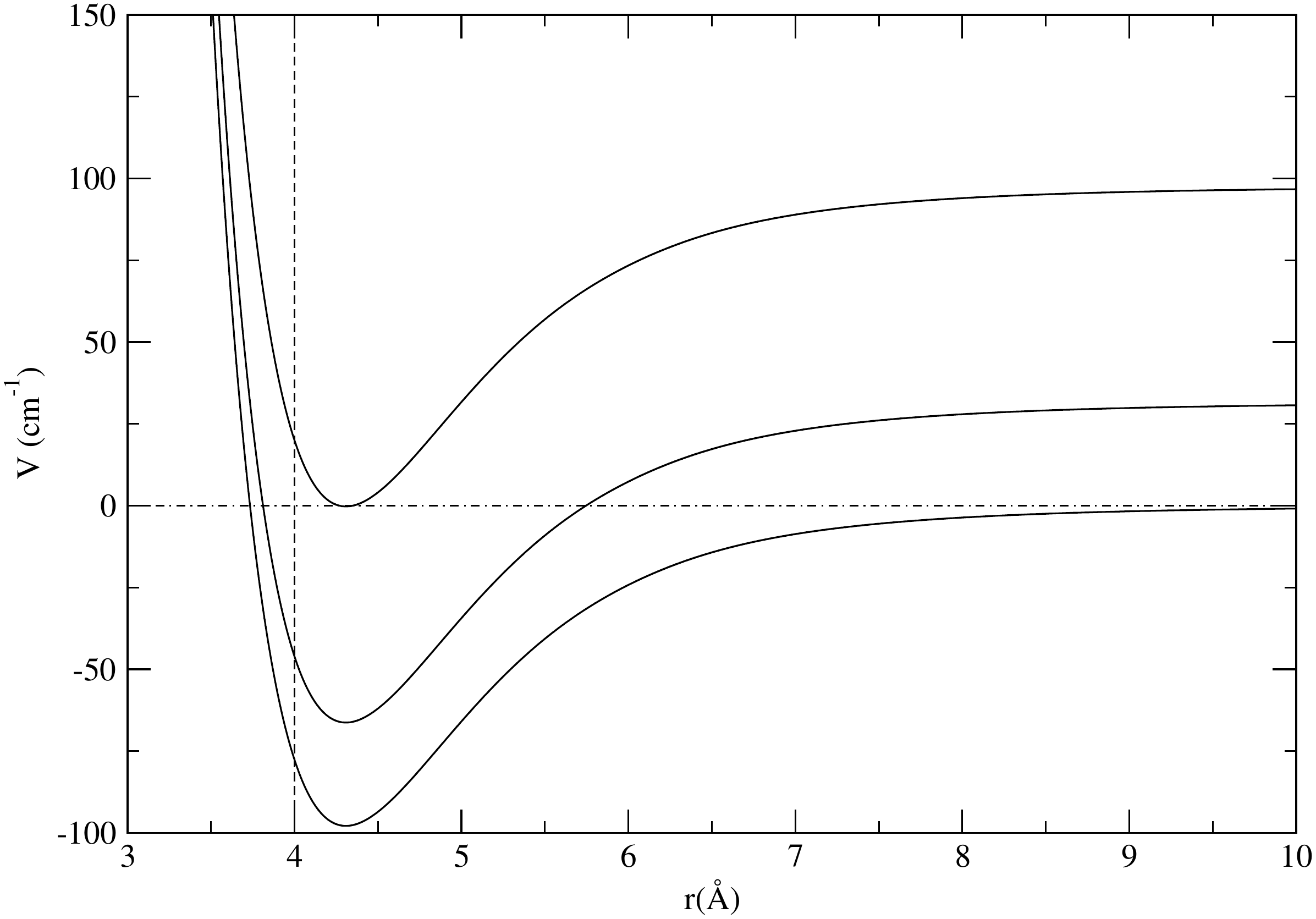}
\caption{The $V_0$ reference potentials for Mg + NH. The first and
second rotational excited state are also shown ($n = 1,2$). The hard wall
at $r = 4.0$~\AA\ is shown as a vertical dashed line. The dot-dashed
horizontal line corresponds to zero energy.}
\label{fig:pot}
\end{figure}

The reference potential contains a hard wall at $r = r^{\text{wall}}_i$, so
that $U^\text{ref}_i(r) = \infty$ for $r<r^{\text{wall}}_i$. In the present
paper we take $r^{\text{wall}}_i = 4.0$~\AA. Figure \ref{fig:pot} shows the
reference potentials for the lowest three rotational states.
All channels with $n \ge 2$ were treated as strongly closed and thus
not included in the MQDT part of the calculation, but were included in
the log-derivative propagation.

\section{Results and discussion}

\begin{figure}[tbh]
\centering
\includegraphics[width=0.86\columnwidth]{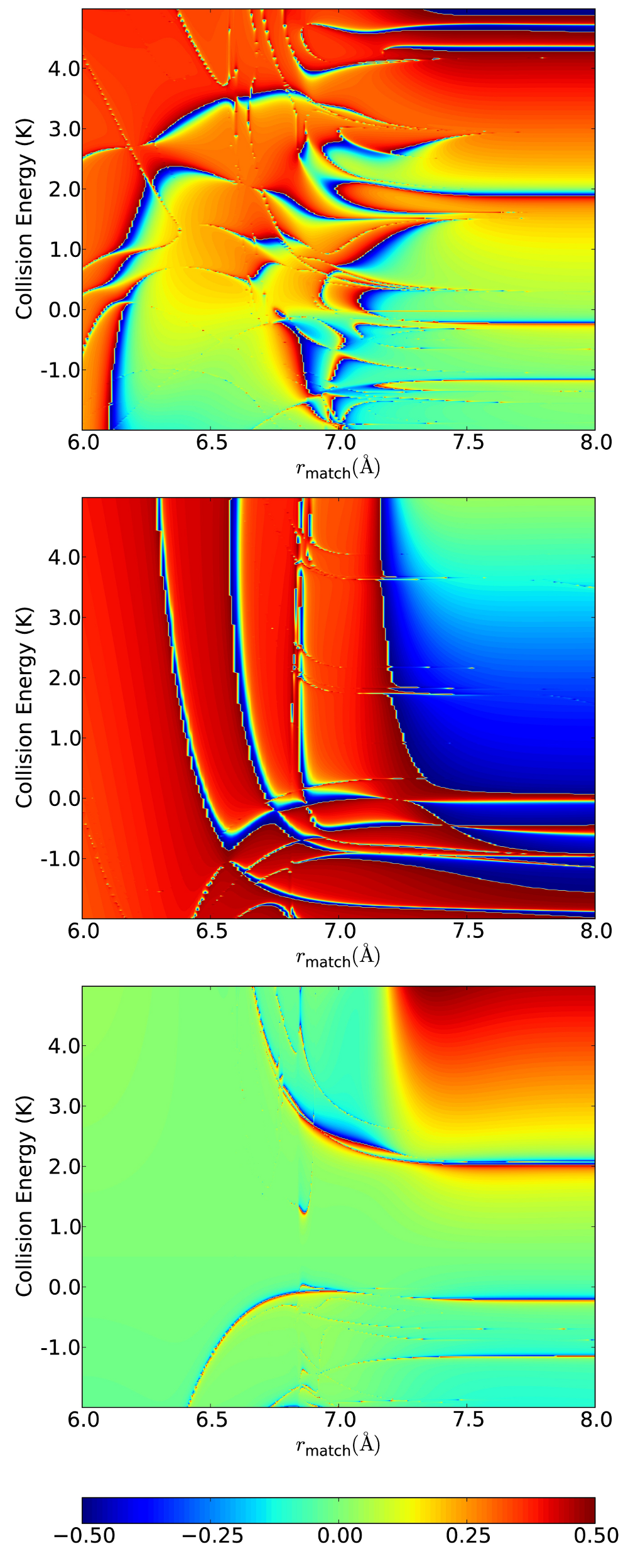}
\caption{(Color online) Contour plot of $\arctan Y_{ii} / \pi$ for a
representative diagonal $\bm Y$ matrix element, $Y_{-1, 8, +2}$, as a function
of energy and $r_\text{match}$ at $B=10$~G. Top panel: obtained with unrotated
reference functions ($\theta_i = 0$). Center panel: obtained with reference
functions rotated by $\theta_i = \pi / 2$. Bottom panel: obtained with
optimized reference functions with $\theta_i = \theta_i^{\rm{opt}}$ in all
channels. The arctangent is show for clarity of plotting: it maps the real
numbers, $\mathbb{R}$, to the domain $-\pi/2$ to $\pi/2$, thus allowing all
magnitudes of $\bm Y$ matrix elements to be seen on a single plot. }
\label{fig:Y_CONTOUR}
\end{figure}

The top panel of Fig.\ \ref{fig:Y_CONTOUR} shows a single diagonal element of
the $\bm{Y}$ matrix, $Y_{-1, 8 +3}$, as a function of the matching distance and
energy, obtained with unrotated reference functions. $Y_{-1, 8, +3}$ is a
representative element of $\bm Y$ with poles at the same locations as the other
elements, chosen to give a good visual representation of the pole structure.
There are many poles visible, which prevent polynomial interpolation over
energies of more that 0.5~K for any value of $r_{\rm match}$ (and much less
than this for some choices of $r_{\rm match}$). The energies of the poles
become independent of $r_{\rm match}$ at long range.

The presence of low-energy poles in $\bm Y$ for some values of $r_{\rm match}$
is a serious problem. For MQDT to be efficient, $r_{\rm match}$ must be chosen
{\em without} solving the coupled equations at many different energies. The
calculations needed to produce contour plots such as those in Fig.\
\ref{fig:Y_CONTOUR} are feasible for a test case such as Mg+NH, but would be
prohibitively expensive for a very large system.

The center panel of Fig.\ \ref{fig:Y_CONTOUR} shows the same element of the
$\bm{Y}$ matrix as a function of the matching distance and energy for reference
functions rotated by $\theta_i = \pi/2$. The poles are in quite different
places, but once again there are many of them. The combination of the top and
center panels demonstrates that, for any arbitrary choice of rotation angle,
poles will appear in the $\bm Y$ matrix, preventing simple interpolation for
most choices of $r_{\rm match}$. This will be true in any MQDT problem with a
large density of resonances. The contour plots do however show that the
position of poles is strongly dependent on the rotation angle, even at large
values of $r_{\rm match}$. This suggests that it will be possible to optimize
the rotation angle in order to move the poles away from the energy range of
interest. It is emphasized that the $\bm S$ matrices obtained from the $\bm Y$
matrices shown in the different panels of Fig.\ \ref{fig:Y_CONTOUR} are
identical.

We now consider how to rotate the reference functions to maximize the pole-free
range over which $\bm Y$ can be interpolated. $Y_{ii}$ as a function of
$\theta_i$ is given by
\begin{equation}
\label{eqn:tan}
Y_{ii} = \tan(\theta_i + \delta_i),
\end{equation}
where $\delta_i$ is the phase shift between the unrotated reference function
$f_i$ and the propagated multichannel wavefunction in channel $i$. There is a
pole in $Y_{ii}$ when $\theta_i +\delta_i = \pi$ and a zero when $\theta_i
+\delta_i = 0$. We thus set $\theta_i^{\rm opt} = -\delta_i$ at one choice of
$r_{\rm match}$, $E$ and $B$, so that the propagated multichannel wavefunction
and the reference wavefunctions are almost in phase and the resulting $\bm Y$
matrix in that region is pole-free.

Because the channels are coupled, rotating the reference functions in one
channel affects the other elements of the $\bm Y$ matrix. In this work we loop
over the channels sequentially, setting each diagonal element to 0 in turn. By
repeatedly looping over all channels, all the diagonal $\bm Y$ matrix elements
are set to 0. For Mg+NH it was sufficient to loop over the channels twice. In a
more strongly coupled system it is expected that this would need to be repeated
more times. This approach allows a set of optimized $\theta_i$ to be obtained
from a single multichannel propagation.

Rotated reference functions have previously been used to transform $\bm Y$
matrices in the study of atomic spectra \cite{Giusti:1984:I, Giusti:1984:II,
Giusti:1984:III, Cooke:1986, Eissner:1969} and atomic collisions
\cite{Osseni:2009}. Adjusting $\theta_i$ at each energy such that $Y_{ii} = 0$
was shown to produce a weak energy dependence of off-diagonal $\bm Y$ matrix
elements across thresholds \cite{Osseni:2009}. However, this approach required
propagating the full multichannel wavefunction many times at different
energies, which is precisely what the present work tries to avoid.

The bottom panel of Fig.\ \ref{fig:Y_CONTOUR} shows how the representative
element $Y_{-1,8,+2}^{\rm opt}$ varies as a function of the matching distance
and energy. All the $\theta_i$ values are optimized as described above at
$E=0.5$~K and $B=10$~G for each value of $r_{\rm match}$, but are {\em not}
reoptimised at each energy. Comparison of this with the upper two panels shows
the effectiveness of optimizing the reference functions. Without optimization,
there were no choices of $r_{\rm match}$ for which $\bm Y$ was pole-free and
thus suitable for interpolation over the energy range of interest. After
optimization, $\bm Y^{\rm opt}$ is pole-free over a substantial range, of about
1~K, for any choice of $r_{\rm match} < 8$~\AA. For values of $r_{\rm match} <
6.5$~\AA, $\bm{Y}^{\rm opt}$ is pole-free over many Kelvin. Beyond 6.5~\AA,
poles start to enter $\bm{Y}^{\rm opt}$ in the energy range of interest. Once
the poles have settled at their asymptotic values at $r_{\rm match} > 7.5$~\AA,
we find that positive energies up to about 2~K are pole-free. However, at
larger values of $r_{\rm match}$ the linearity of $\bm{Y}^{\rm opt}$ over the
pole-free region decreases. This is due to negative energy poles in the $\bm Y$
matrix which our procedure cannot move significantly. There is one particularly
bad choice of $r_{\rm match}$ at $\approx 6.8$~\AA, but provided this unlucky
choice of $r_{\rm match}$ is avoided, $\bm{Y}^{\rm opt}$ can be interpolated
smoothly over the positive energy range from 0 to $>2$~K for any choice of
$r_{\rm match}$.

\begin{figure}[tb] \centering
\includegraphics[width=1\columnwidth]{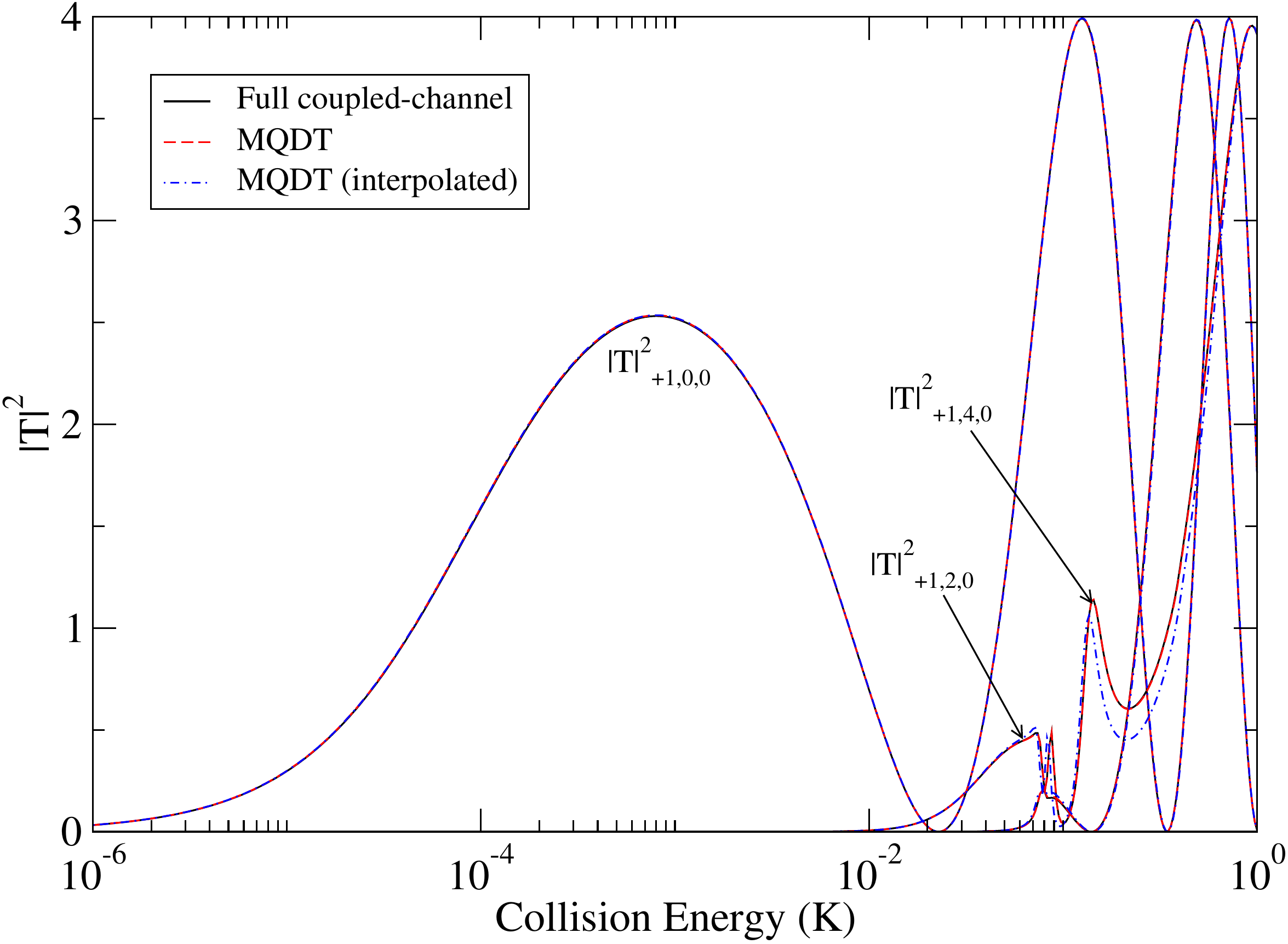}
\caption{(Color online) The squares of diagonal T-matrix elements
$T_{m_j,L,M_L}$ in the incoming channels for $m_j = +1$ and $L = 0$, 2
and 4 at $B = 10$ G, obtained from full coupled-channel calculations
(solid, black) and MQDT with optimized reference functions for
$r_{\rm match}$ = 6.5 \AA, both with (dot-dash, blue) and without
(dashed, red) interpolation.} \label{fig:S}
\end{figure}

Figure \ref{fig:S} compares diagonal T-matrix elements $|T_{ii}|^2$ (where
$T_{ij}=\delta_{ij}-S_{ij}$) obtained from full coupled-channel calculations
with those from the MQDT method, with a matching distance of
$r_\text{match}=6.5$~\AA, using reference functions optimized at 0.5~K. MQDT
results were obtained both by recalculating the $\bm{Y}$ matrix at every energy
and by interpolating $\bm{Y}^{\rm opt}$ linearly between two points separated
by 1~K. The MQDT results with $\bm Y$ recalculated at each energy can scarcely
be distinguished from the full coupled-channel results. The MQDT results
obtained by interpolation are also very similar to the full coupled-channel
results except around the resonance feature at $E \approx 0.1$~K. The
interpolated result could of course be improved simply by performing
coupled-channel calculations to obtain $\bm{Y}^{\rm opt}$ at one or two extra
energies across the range, to allow for a higher-order interpolation, or by
using a linear interpolation over a smaller energy range.

\begin{figure}[tb]
\centering
\includegraphics[width=1\columnwidth]{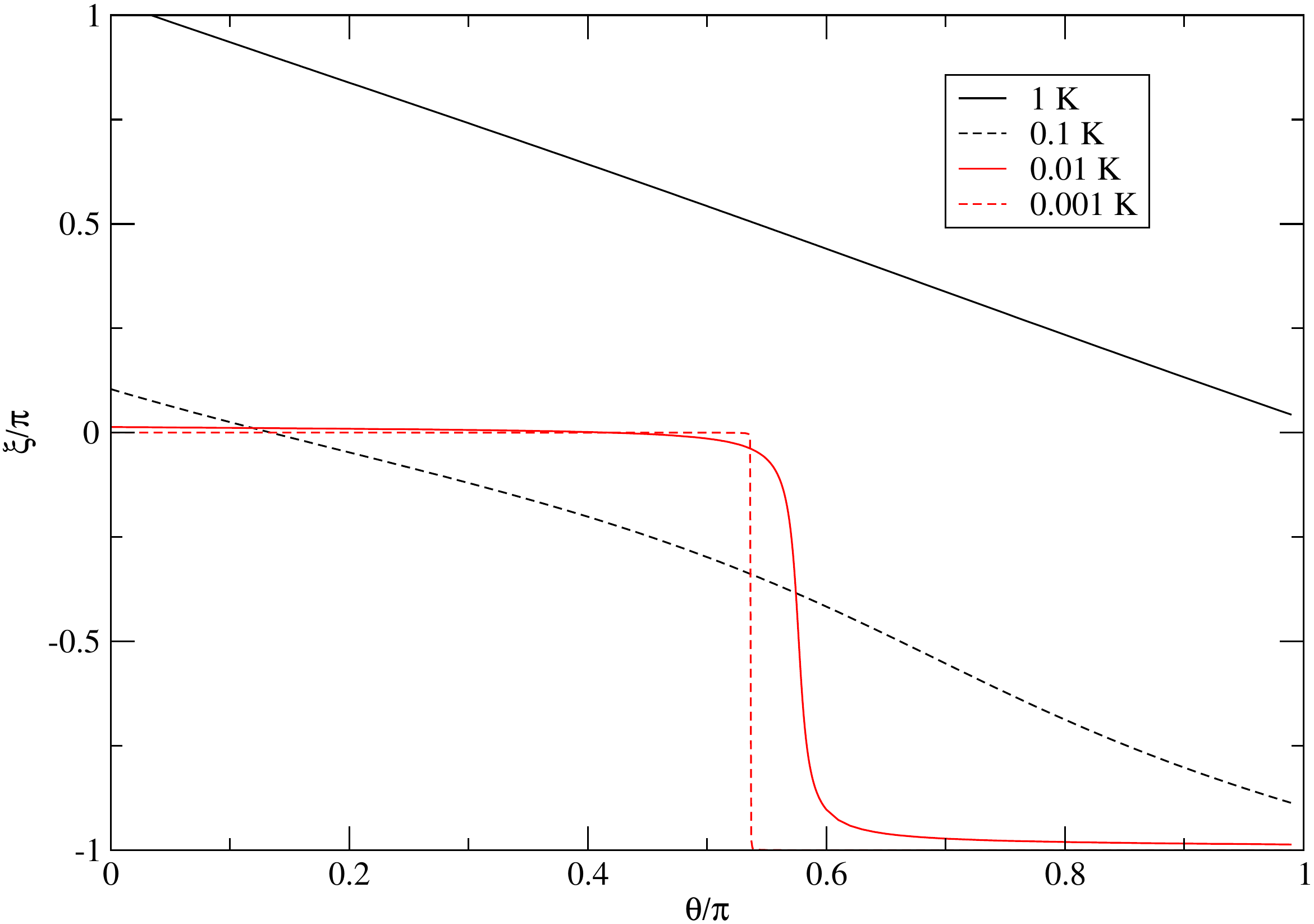}
\caption{(Colour online) The asymptotic phase shift $\bar\xi_i$ as a function
of the rotation angle $\theta$ for the incoming d-wave channel ($+1,2,0$).}
\label{fig:xi}
\end{figure}

In this work we use $\theta_i$ to rotate our short-range reference functions
$f_i$ and $g_i$. In principle, we could rotate the reference functions by
varying the asymptotic phase shifts $\xi_i$ instead of the short-range phases
$\theta_i$. However Figure \ref{fig:xi} shows why this is not desirable.
Due to the highly nonlinear relationship between $\xi_i$ and $\theta_i$,
obtaining the optimum rotation angle of the short-range reference functions
$f_i$ and $g_i$ by varying the angle $\xi_i$ would be laborious at very low
collision energies.

\subsection{Magnetically tunable Feshbach resonances}

The effects of magnetic fields on cold molecular collisions are very
important, since collisions can be controlled by taking advantage of
magnetically tunable low-energy Feshbach resonances. We are therefore
interested in how $\bm S$ matrix elements behave as a function of
magnetic field across Feshbach resonances. It is thus important that
the $\bm Y$ matrix is weakly dependent on magnetic field in such
regions.

\begin{figure}[tb]
\centering
\includegraphics[width=1\columnwidth]{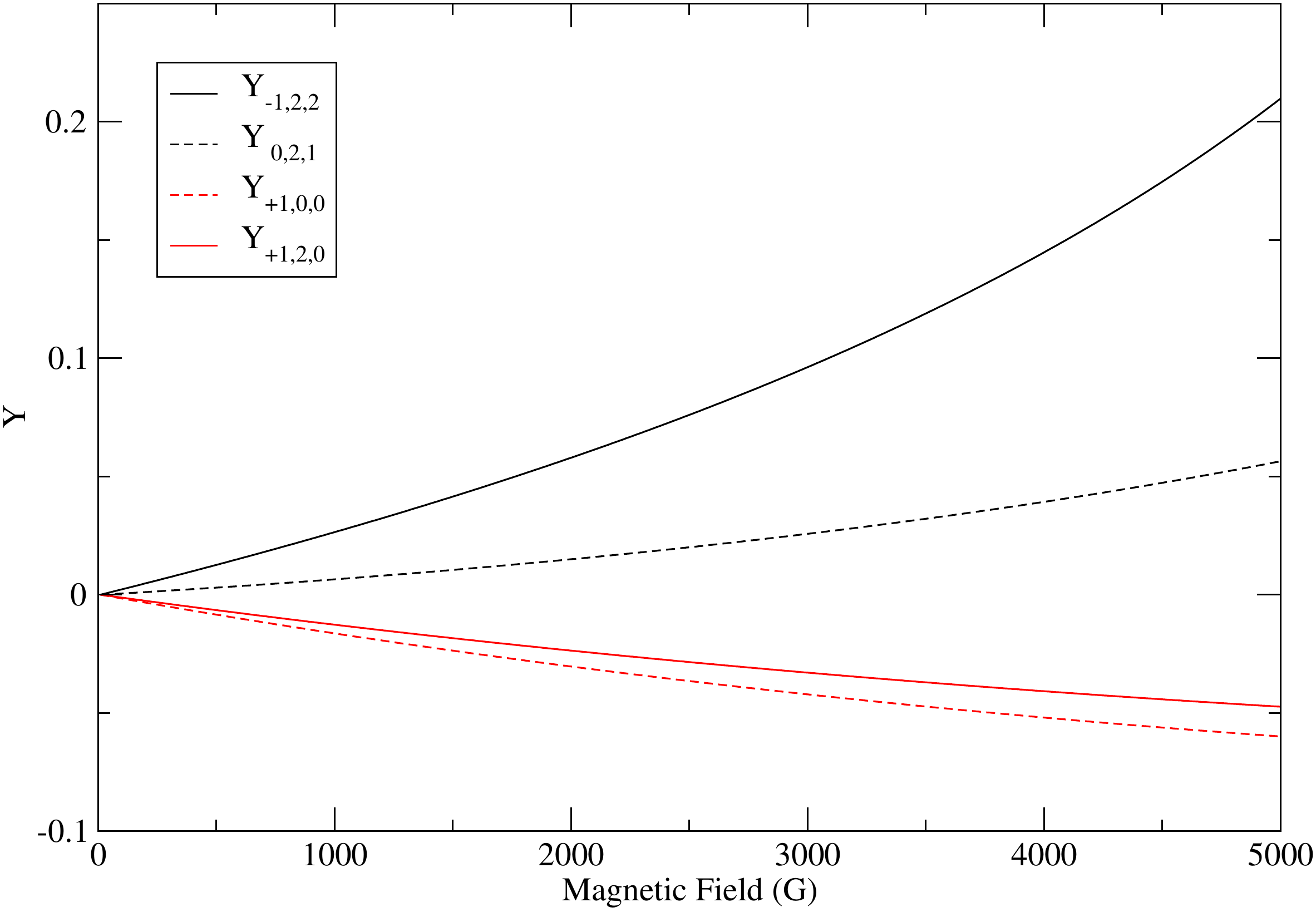}
\caption{(Color online) Representative $\bm Y^{\rm{opt}}$ matrix
elements as a function of field at $E=1$~mK.} \label{fig:Y_FIELD}
\end{figure}

Figure \ref{fig:Y_FIELD} shows the diagonal elements of the optimized $\bm Y$
matrix as a function of magnetic field for Mg + NH collisions over the range
from 10~G to 5000~G for a collision energy of 1~mK. This range of fields tunes
across 6 Feshbach resonances. The reference functions were optimized at 10~G
and 1~mK. The elements of $\bm{Y}^{\rm opt}$ are smoothly curved over the
entire 5000~G range and could be well represented by a low-order polynomial.

\begin{figure}[tb]
\centering
\includegraphics[width=1\columnwidth]{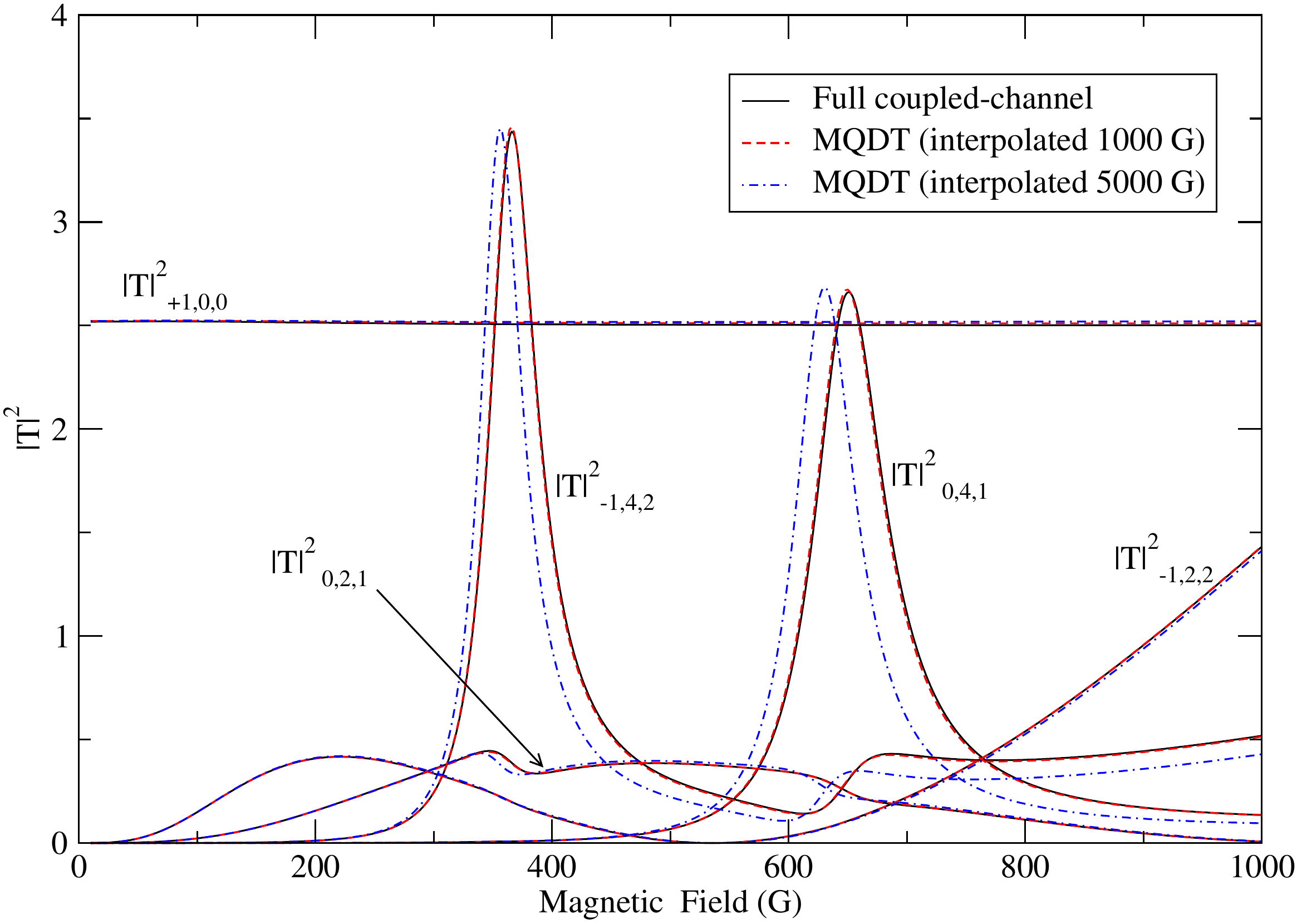}
\includegraphics[width=1\columnwidth]{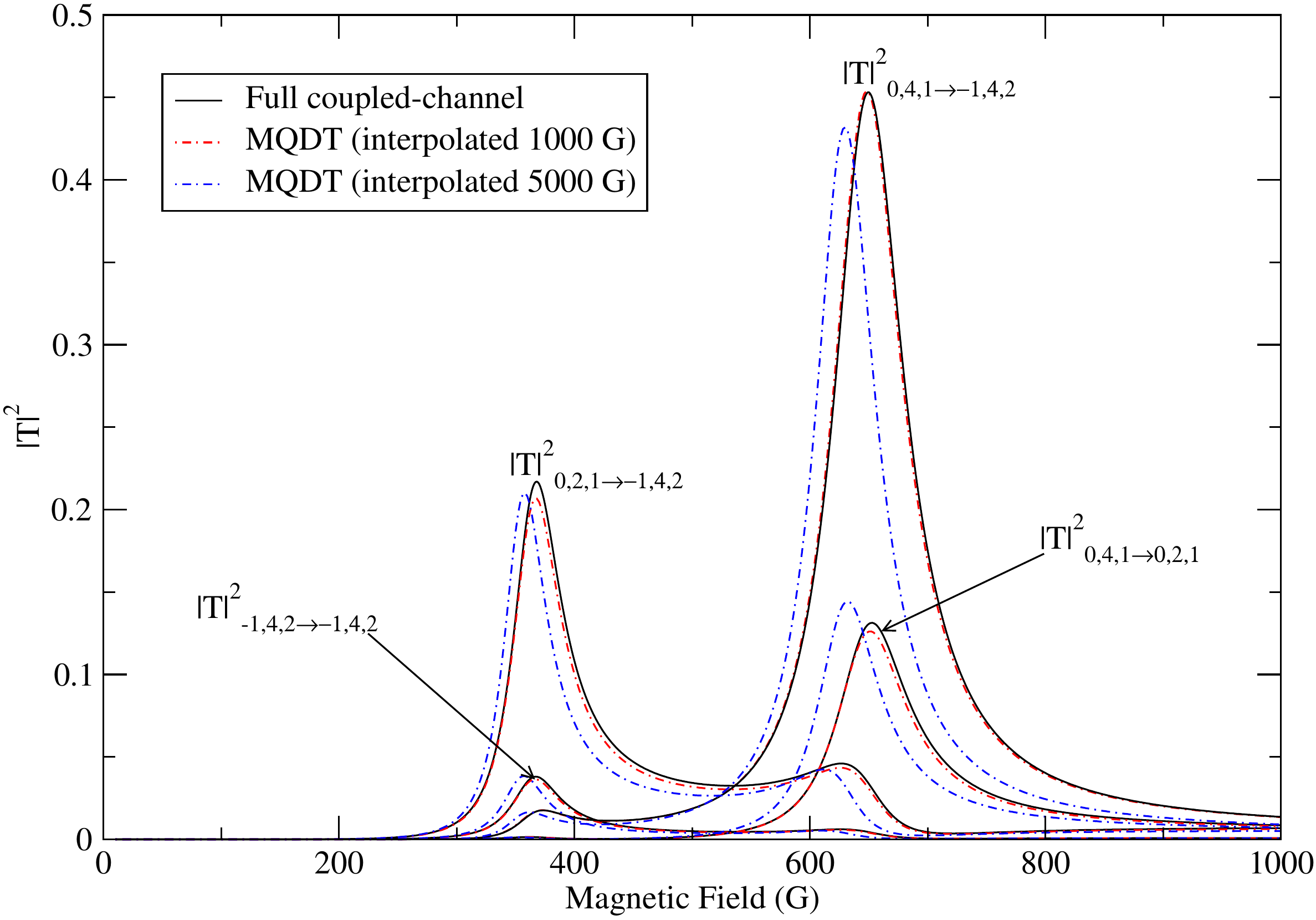}
\caption{(Color online) Squares of T-matrix elements at 1~mK as a function of
field in the vicinity of a Feshbach resonance. Upper panel: diagonal elements;
lower panel: off-diagonal elements.} \label{fig:S_FIELD}
\end{figure}

Figure \ref{fig:S_FIELD} shows the comparison between optimized MQDT and full
coupled-channel calculations for a selection of diagonal and off-diagonal
T-matrix elements as the magnetic field is tuned at 1~mK. The reference
functions were optimized at 10~G and 1~mK and MQDT results were obtained by
linear interpolation of $\bm{Y}^{\rm opt}$ between two points separated by
1000~G and by 5000~G. Interpolation over 1000~G gives resonance features that
are in very good agreement with the full coupled-channel calculation to better
than 1~G. Interpolation over 5000~G gives resonance features of the correct
shape, with positions that are still within about 10~G of the full
coupled-channel results. The difference between the interpolated result and the
full coupled-channel calculation is a result of both the choice of $r_{\rm
match}$ and the interpolation. The quality of the interpolation could be
improved by considering a few more fields across the range to allow for
higher-order polynomial interpolation or by using linear interpolation over a
smaller field range.

Full MQDT calculations recalculating the $\bm Y$ matrix at every magnetic field
give resonance positions accurate to 0.4~G. The remaining errors between the
full coupled-channel calculations and the MQDT results will reduce with a
larger choice of $r_{\rm match}$. As seen in the bottom panel of Fig.\
\ref{fig:Y_CONTOUR}, the optimized $\bm Y$ matrices obtained at larger values
of $r_{\rm match}$ are still amenable to interpolation, though over a more
restricted energy range.

\section{Conclusions}

We have shown that Multichannel Quantum Defect Theory can provide an efficient
computational method for low-energy molecular collisions as a function of both
energy and magnetic field. In particular, we have shown how a disposable
parameter of MQDT, the phase of the short-range reference functions, may be
chosen to make the MQDT $\bm Y$ matrix smooth and pole-free over a wide range
of energy and field. This smooth variation allows the $\bm Y$ matrix to be
evaluated from coupled-channel calculations at a few values of the energy and
field and then to be obtained by interpolation at intermediate values. It is
not necessary to repeat the expensive coupled-channel part of the calculation
on a fine grid.

The procedure developed here is to choose the phase of the reference functions
in each channel so that the diagonal $\bm Y$ matrix in each channel is zero at
a reference energy and field. This ensures that there are no poles in the $\bm
Y$ matrix, which would prevent smooth interpolation, close to the reference
energy. Optimizing the phase in this way is very inexpensive, and once it is
done the cost of calculations at additional energies and fields varies only
linearly with the number of channels $N$, not as $N^3$ as for full
coupled-channel calculations. MQDT with optimized $\bm Y$ matrices is thus a
very promising alternative to full coupled-channel calculations for cold
molecular collisions, particularly when fine scans over collision energy and
magnetic field are required.

The $\bm Y$ matrix is defined to encapsulate all the collision dynamics that
occurs inside a matching distance $r_{\rm match}$, and the choice of this
distance is important. There is a trade-off between the accuracy of the method
and the size of the pole-free region of the optimized $\bm Y$ matrix. For large
values of $r_{\rm match}$, resonant features may appear in the $\bm Y$ matrix
and prevent simple interpolation over large ranges of energy and field. For
smaller values of $r_{\rm match}$, optimizing the reference functions allows
interpolation over many Kelvin, but the accuracy of MQDT is reduced because
interchannel coupling is neglected outside $r_{\rm match}$.

For the moderately anisotropic Mg + NH system studied here, optimized MQDT with
an interpolated $\bm Y$ matrix can provide numerical results in quantitative
agreement with fully converged coupled-channel calculations. In future work, we
will investigate the extension of this approach to more strongly coupled
systems, with larger anisotropy of the interaction potential and more closed
channels that produce scattering resonances.

\section{Acknowledgments}

JFEC is grateful to EPSRC for a High-End Computing Studentship. The authors are
grateful for support from EPSRC, AFOSR MURI Grant FA9550-09-1-0617, and EOARD
Grant FA8655-10-1-3033

\bibliography{all,james}

\end{document}